\documentstyle[aps,twocolumn,10pt]{revtex}
\begin{document}
\twocolumn[\hsize\textwidth\columnwidth\hsize\csname
@twocolumnfalse\endcsname
\title{Mean Field Dynamics in Non-Abelian
Plasmas from Classical Transport Theory}
\author{
\hfill 
Daniel F. Litim$^a$ and Cristina Manuel$^b$
\hfill\raisebox{21mm}[0mm][0mm]{\makebox[0mm][r]{ECM-UB-PF-99-04, 
CERN-TH-99-29, February 1999}}%
}
\address{${}^a$Departament ECM, Facultat de 
$F\!\raisebox{.15ex}{\'}\!\!\!\,{\i}sica$ {\sl \&} IFAE, 
Universitat de Barcelona\\
Diagonal 647, E 08028 Barcelona, Spain. E-Mail: Litim@ecm.ub.es\\
${}^b$Theory Division, CERN, CH-1211 Geneva 23, Switzerland.
E-Mail: Cristina.Manuel@cern.ch}
 
\maketitle

\begin{abstract}\noindent
{Based on classical transport theory, we present a general set of 
covariant equations describing the dynamics of mean fields and their 
statistical fluctuations in a non-Abelian plasma in or out-of-equilibrium. 
A procedure to obtain the collision integrals for the Boltzmann 
equation from the microscopic theory is described.
As an application, we study a hot non-Abelian plasma close to equilibrium,
where the fluctuations are integrated out explicitly. 
For soft fields, and at logarithmic accuracy, we obtain B\"odeker's 
effective theory.\\[1ex]
}
\end{abstract}
\vskip1.5pc]

The dynamics of mean fields in non-Abelian plasmas is at the basis for 
an understanding of many properties of the early universe.
This concerns the quark-gluon plasma which 
is believed to be formed at high temperature and/or large chemical 
potential, or the plasma formed in the symmetric phase of the electroweak 
theory.  The energy densities required to form a quark-gluon plasma might  
be reached in the coming generation of heavy ion 
colliders.  A theoretical framework
to describe the dynamics in non-Abelian plasmas in or out of equilibrium
is thus mandatory (see \cite{ElzeHeinz} for an early review).

The dynamics of soft fields with 
momenta about $g^2 T$ is dominated by non-perturbative 
phenomena, even for small gauge coupling $g\ll  1$ and close to 
equilibrium \cite{Gross}. Here, some progress has been 
achieved recently by B\"odeker \cite{Bodeker}, who derived an effective 
theory for the soft modes after integrating-out the perturbative physical 
scales $T$ and $gT$ from the field theory. 

In this Letter, we study mean field dynamics in the context of classical 
transport theory \cite {Heinz}. In the close-to-equilibrium plasma, classical 
transport theory is known to describe
correctly the physics around the scale $gT$ \cite{KLLM}, in full agreement
with the results obtained from quantum field theory \cite{HTL}. 
We will show how also the physics at scales $g^2 T$, 
and at logarithmic accuracy, can be understood in the same kinetic language. 
Our ultimate goal is providing a well-defined prescription to treat a number 
of interesting problems including out-of-equilibrium 
phenomena. Here, we present the main results 
of our analysis. This includes the derivation of a general set of dynamical 
equations for the mean fields and their fluctuations, and a prescription
for obtaining collision integrals for Boltzmann equations. An explicit
application for hot non-Abelian plasmas close to equilibrium is given as well.
We leave for a longer publication a more extensive discussion and further 
details on the computations \cite{future}.

We begin with the derivation of a set of equations that describe both 
the dynamics of fluctuations and of the mean fields in a non-Abelian plasma. 
We follow the same philosophy and methods which have been  used longly in 
kinetic theory to study Abelian plasmas (see \cite{K}). Our equations 
can be seen as the generalization to non-Abelian plasmas. 

The starting point is to consider gluons (or electrons, quarks) 
as charged point 
particles moving on a world line according to their classical equations of 
motion. They interact via a classical chromo-electromagnetic field. We 
consider a system of these particles carrying a non-Abelian color charge 
$Q^a$, where the color index runs from $a=1$ to $N^2-1$ for a SU($N$) gauge 
group. 

Within a microscopic description, the trajectories in phase space, and 
therefore the one-particle phase space density $f(x,p,Q)$, 
are known exactly. The classical
trajectories $x(\tau), p(\tau)$ and $Q(\tau)$ 
are solutions of the Wong equations \cite{Wong}
\begin{eqnarray}
&&\displaystyle m\frac{dx^\mu}{d\tau}=p^\mu,\quad m\frac{dp^\mu}{d\tau}=g\, 
Q^aF_a^{\mu\nu} p_\nu \ , \nonumber 
\\ &&\displaystyle m\frac{dQ^a}{d\tau}=-g\, p_\mu f^{abc}A_b^\mu\,Q_c \ .
\label{Wong}
\end{eqnarray}
Here, $A_\mu^a $ denotes the microscopic vector gauge field, 
$F^{a}_{\mu\nu}[A]=\partial_\mu A^{a}_\nu-\partial_\nu A^{a}_\mu
+ g f_{abc}A^b_\mu A^c_\nu$ the corresponding
microscopic field strength, and $f^{abc}$ are the structure constants of 
SU($N$). We set
$c=\hbar=k_B =1$, and work in natural units. Effects of spin 
will be neglected. 

The microscopic phase space density obeys Liouville's 
theorem $df/d\tau=0$ \cite{KLLM}.
We can write it using (\ref{Wong}) as \cite{Heinz}
\begin{mathletters}\label{NA-Micro}
\begin{equation}
p^\mu\left[\partial_\mu-g f^{abc} A_\mu^b Q_c {\partial^Q_a}
- gQ_aF^a_{\mu\nu}{\partial_p^\nu}
\right]f=0\ ,\label{NA-f}
\end{equation}
with $\partial^p_\mu\equiv{\partial }/{\partial  p^\mu}$ 
and $\partial ^Q_a\equiv{\partial }/{\partial  Q^a}$. 
This equation is gauge invariant, with $f$ transforming as a scalar 
\cite{KLLM}, and is completed with the Yang-Mills equation,
\begin{equation}\label{NA-J}
D_\mu F^{\mu\nu}(x) = J^\nu(x)\ .
\end{equation}
In the adjoint, the covariant derivative is given by
$D^{ac}_\mu[A]=\partial_\mu\delta^{ac}+gf^{abc} A_\mu^b$. 
In a self-consistent picture the current $J^\mu_a(x)$ is due to the 
particles themselves, thus 
\begin{equation}\label{current}
J^\mu_a(x)=g \sum_{\hbox{\tiny helicities}\atop\hbox{\tiny species}}
\int dPdQ \ p^\mu\, Q_a\, f(x,p,Q) \ . 
\end{equation}
\end{mathletters}%
(Throughout, we shall omit a species or helicity index on $f$, and 
the sum over 
species and helicities will, in the sequel, not be given explicitly.) 
Physical constraints are enforced through 
the phase space volume 
element $dP\equiv d^4 p \ 2 \Theta(p_0)\delta(p^2-m^2)/(2\pi)^3$, while
 $dQ$ contains 
$\delta$-functions imposing the
group Casimirs (see \cite{KLLM} for their definition). The covariant 
conservation of the current (\ref{current}) is shown using 
(\ref{NA-f}) \cite{KLLM}.

If the system under study contains a large number of particles it is 
impossible to follow their individual trajectories in phase space. Thus, 
$f$ can no longer be considered a deterministic quantity and one has to 
switch to a statistical description, taking statistical averages 
$\langle \ldots \rangle$ of all microscopic quantities. We write
\begin{equation}
A^{a}_\mu = {\bar A}^a_\mu + a^a_\mu \ ,\  
f = {\bar f} + \delta f \ ,\ J^\mu_a=\bar J^\mu_a +\delta J^\mu_a \ , 
\label{NA-delta-a} 
\end{equation}
where the quantities with a bar denote the mean values, {\it e.g.} 
$\bar f = \langle f \rangle$ and $\bar A = \langle A \rangle$, 
while the mean values of fluctuations 
vanish, $\langle \delta f \rangle=0$ and $\langle a \rangle=0$. 
We also split
\begin{mathletters}
\begin{eqnarray}
&&F^{a}_{\mu\nu} = {\bar F}^a_{\mu\nu} + f^a_{\mu\nu}  ,\\
&&f^a_{\mu\nu}=(\bar D_\mu a_\nu-\bar D_\nu a_\mu)^a
+g f^{abc}a^b_\mu a^c_\nu  \ ,
\end{eqnarray}
\end{mathletters}%
with $\bar D\equiv D[\bar A]$ and $\bar F\equiv F[\bar A]$. Note, that 
the mean field strength $\langle F^a_{\mu\nu}\rangle=\bar F^a_{\mu\nu}+
g f^{abc}\langle a^b_\mu a^c_\nu\rangle$ due to the non-Abelian nature
 of the theory.

Let us take a statistical average of (\ref{NA-Micro}) to find the kinetic 
equations for the mean values,
\begin{mathletters}\label{NA-Macro}
\begin{eqnarray}
&&p^\mu\left(\bar D_\mu- gQ_a\bar F^a_{\mu\nu} 
 \partial _p^\nu\right)\bar f=\left\langle\eta\right\rangle
+ \left\langle\xi\right\rangle \ , \label{NA-1}\\
&&\bar D_\mu \bar F^{\mu\nu}  + \left\langle J_{\mbox{\tiny fluc}}^{\nu}
\right\rangle =\bar J^\nu \ .\label{NAJ-1} 
\end{eqnarray}
\end{mathletters}%
In (\ref{NA-1}) we used 
$[\partial _\mu-g f^{abc}Q_c A^b_\mu{\partial ^Q_a}]f\equiv D_\mu f$.
The functions $\eta, \xi$ and $J_{{\mbox{\tiny fluc}}}$ are of 
second and higher order 
in the fluctuations and read 
\begin{mathletters}\label{NA-func}
\begin{eqnarray}
\eta\ &\equiv& g Q_a\, p^\mu \partial _p^\nu   f_{\mu\nu}^a\, 
\delta f  \ , \label{NA-eta}\\
\xi\ &\equiv& gp^\mu f^{abc}Q^c\left(\partial ^Q_a a_\mu^b\ 
\delta f\ + g a_\mu^a a_\nu^b\partial^\nu_p\bar f\right) ,\label{NA-xi}\\
J_{{\mbox{\tiny fluc}}}^{a,\nu}&\equiv& g f^{dbc} \left[\bar D^{\mu}_{ad}  
a_{b,\mu} 
a_c^\nu  + \delta_{ad} a_{b,\mu}\, f_c^{\mu\nu} \right] .\label{NA-Jfluc}
\end{eqnarray}
\end{mathletters}%
The corresponding equations for the fluctuations are obtained by  
subtracting (\ref{NA-Macro}) from (\ref{NA-Micro}). The result is
\begin{mathletters}\label{NA-fluc}
\begin{eqnarray}
&&p^\mu\left[\bar D_\mu - gQ_a\bar F^a_{\mu\nu}\partial _p^\nu\right]
\delta f- gp^\mu a_{b,\mu} f^{abc} Q_c\partial ^Q_a \bar f\nonumber\\
&&\ - g Q_a[\bar D_\mu a_\nu-\bar D_\nu a_\mu]^a p^\mu \partial ^p_\nu\bar f
=\eta  + \xi - \left\langle\eta+\xi\right\rangle\label{NA-2}\\
&&\left[\bar D^2 a^\mu-\bar D^\mu(\bar D_\nu a^\nu)\right]^a+2 gf^{abc}
\bar F_b^{\mu\nu}a_{c,\nu}+\nonumber\\&&\qquad\qquad\qquad \qquad\qquad 
+J_{{\mbox{\tiny fluc}}}^{a,\mu}-\left\langle J_{{\mbox{\tiny fluc}}}^{a,\mu}\right\rangle
=\delta J^{a,\mu} \ .\label{NAJ-2}
\end{eqnarray}
\end{mathletters}%
A number of comments are in order.

\begin{list}{}{\leftmargin=.4cm\itemsep=0.05cm}

\item[1.] The equations (\ref{NA-Macro}) and (\ref{NA-fluc}) 
are exact, no approximations have been made. 
In particular, they are also valid in out-of-equilibrium situations.

\item[2.] The equations (\ref{NA-Macro}) and (\ref{NA-fluc}) 
are consistent with gauge invariance. They are covariant under the
mean gauge field symmetry   
$\delta_\epsilon \bar A^a_\mu = (\bar D_\mu \epsilon)^a$ and 
$\delta_\epsilon a_\mu^a      = g f^{abc}a^b_\mu \epsilon^c$, in 
analogy to the background field formalism \cite{BGS}. 
This establishes the
compatibility of the statistical average with the gauge transformations  
of the mean field. We postpone a careful 
and detailed discussion to \cite{future}. 

\item[3.] The microscopic current conservation implies 
\begin{equation}\label{currentconservation}
\left(\bar D_\mu \bar J^\mu\right)_a + g f_{abc}\left\langle a^b_\mu\, 
\delta J^{c,\mu}\right\rangle = 0\,.
\end{equation}
This is automatically consistent with (\ref{NAJ-1}), provided $\bar J$ and 
$\delta J$ are solutions of (\ref{NA-Macro}) and (\ref{NA-fluc}). (A similar 
equation holds for the fluctuation fields.)  Note that the validity 
of (\ref{currentconservation}) turns into a non-trivial consistency check 
for approximate solutions.

\item[4.] The functions $\langle\eta\rangle$ and $\langle\xi\rangle$  can be 
considered as the effective collision integrals of the Boltzmann
equation (\ref{NA-1}). In our formalism the collision integrals 
arise as correlators of statistical fluctuations. 
The fluctuations of the gauge fields cause random 
changes in the motion of particles, and thus, they can be viewed 
as having the same effects as collisions. This can be seen as a derivation of
 collision integrals from the microscopic theory.
Note also that the current induced by the fluctuations of the gauge field 
$\langle J_{{\mbox{\tiny fluc}}}\rangle$ is a purely non-Abelian effect.

\item[5.] A general procedure for integrating-out the fluctuations 
amounts to first solve  their dynamics (\ref{NA-fluc})
in the background of mean fields. In general, this is a 
difficult task, in particular due to the non-linear terms  in (\ref{NA-fluc}).
The (explicit) solution is then inserted into (\ref{NA-func}). The 
back-coupling of the fluctuations to the mean fields is finally obtained 
after taking the statistical average of the functions (\ref{NA-func}), and 
yields the effective collision integrals and the induced current in 
(\ref{NA-Macro}). 

\item[6.] The set of equations (\ref{NA-Macro}) and (\ref{NA-fluc}) reproduces 
the known set of kinetic 
equations for Abelian plasmas in the corresponding limit \cite{K}, 
in which only the collision 
integral $\langle\eta\rangle$ survives. The Abelian 
counterpart of $\langle\eta\rangle$ can be
expressed as the Balescu-Lenard collision integral \cite{K}.  
One can then proof in a rigorous way the
correspondence between fluctuations and collisions in the 
Abelian plasmas mentioned  above. 
(An analogous
derivation of collision integrals for Wigner functions can be found in
\cite{Selikhov}, see also \cite{SG}.) 

\item[7.] Neglecting all fluctuations reduces (\ref{NA-Macro}) 
to the well-known (non-Abelian) Vlasov equations. 
\end{list}

This terminates the derivation and discussion of the basic set of equations.\\
 
To put the method to work we will specialize our analysis to hot non-Abelian 
plasmas close to equilibrium, with the gauge coupling $g\ll 1$. 
This allows us to perform several approximations. 
We will consider small fluctuations, neglecting  in (\ref{NA-eta}) and 
(\ref{NA-Jfluc}) the terms cubic in the fluctuations. This is interpreted as 
neglecting effective three body collisions versus binary ones. 
In the same spirit, we employ the second-moment approximation for the 
dynamics of the fluctuations \cite{K}, setting 
$\eta =\left\langle \eta\right\rangle$, 
$\xi = \left\langle\xi\right\rangle$ and 
$J_{{\mbox{\tiny fluc}}}=\left\langle J_{{\mbox{\tiny fluc}}}\right\rangle$ 
in (\ref{NA-fluc}). This linearizes the dynamics of the fluctuations and 
can be interpreted as neglecting the influence of 
collisions on the dynamics of the fluctuations.  Finally, the term 
containing the mean field strength in (\ref{NA-2}) is negligible compared 
to the remaining terms and will be omitted, as long as 
$g\, |\bar F^{\mu\nu}_a| /m_D\ll T $, with $m_D$ the Debye mass \cite{K}. 

We study the mean distribution function
$\bar f(x,p,Q)=$ $\bar f^{\rm eq.}(p_0)+ g \bar f^{(1)}(x,p,Q)$. 
In the strictly classical approach, the relativistic Maxwell 
distribution at equlibrium is used for any species of particles. 
Here, we consider only massless particles in the adjoint, with
 $\int dQ\, Q_aQ_b=N\,\delta_{ab}$. For particles in the fundamental one 
has $\int dQ\, Q_aQ_b=\mbox{\small{$\frac{1}{2}$}}\,\delta_{ab}$ instead. 
Solving (\ref{NA-Macro}) for vanishing 
fluctuations in this approximation  gives the infinite set of non-Abelian 
hard thermal loops \cite{KLLM,HTL}.  

We now include  small statistical fluctuations $\delta f$ around $\bar f$  
 and re-write the approximations to (\ref{NA-Macro}) and (\ref{NA-fluc}) 
in terms of current densities and their fluctuations. Consider the current 
densities
\begin{mathletters}
\begin{eqnarray}
J_{a_1\cdots a_n}^\rho(x,p)&=&g\ p^\rho  \!\int\! 
dQ Q_{a_1}\cdots Q_{a_n} f(x,p,Q), \label{NA-Jp}\\
{\cal J}_{a_1\cdots a_n}^\rho(x, v)&=&\int d\tilde P J
_{a_1\cdots a_n}^\rho(x,p) \ .\label{NA-Jv}
\end{eqnarray}
\end{mathletters}%
The measure $d\tilde P$ only integrates over the radial 
components, $dP=d\tilde P d\Omega/4\pi$, and
$v^\mu=(1,{\bf v})$ with ${\bf v}^2=1$. The current (\ref{current}) 
is obtained performing the remaining angle integration
$J(x)=\int \mbox{\small{$\frac{d\Omega}{4\pi}$}}{\cal J}(x,v)$. 
From now on we will omit the arguments of the
current density $\cal J$, unless necessary to avoid confusion.

After multiplying (\ref{NA-1}) by $g Q_a p^\rho/p_0$, summing over the two
helicities, 
 and integrating over $d\tilde P dQ$, we obtain for the mean 
current density at leading order in $g$
\begin{mathletters}
\begin{eqnarray} \label{soft-mean}
&&v^\mu \bar D_\mu \bar {\cal J}^\rho+ m^2_D v^\rho v^\mu \bar F_{\mu0}
=\left\langle \eta^\rho\right\rangle +\left\langle \xi^\rho\right\rangle \,,\\
&&\bar D_\mu\bar F^{\mu\nu} 
+\left\langle J_{\mbox{\tiny fluc}}^{\nu}\right\rangle =
 \bar J^\nu \,,\label{soft-YM}
\end{eqnarray}
\end{mathletters}%
with the Debye mass
$m_D^2= - 2 g^2  N \!\int\! d \tilde P p_0\, d \bar f^{\rm eq}(p_0)/d p_0 $, and 
\begin{mathletters}\label{NA-log-1}
\begin{eqnarray}
\eta^\rho_a&=&- g
\int \frac{d \tilde P}{p_0} \Big\{
 (\bar D_\mu a^\rho-\bar D^\rho a_\mu)^b\, \delta J^{\mu}_{ab}(x,p) 
 \nonumber\\ &&\qquad\qquad
-  \frac{p^\rho}{p_0}  (\bar D_\mu a_0-\bar D_0 a_\mu)^b \,
\delta J^{\mu}_{ab}(x,p) \Big\} \,,\label{NA-eta1}\\ 
\xi^\rho_a &=&-g  f_{abc} v^\mu\, a^b_{\mu} \ \delta {\cal J}^{c,\rho} 
\,,\label{NA-xi1}\\
 J_{\mbox{\tiny fluc}}^{\rho,a}&=&  g f^{dbc}\! \left\{\bar D_{\mu}^{ad}  
a_{b}^{\mu} a_c^\rho  
+\delta^{ad}  a^b_\mu\! \left(\bar D^\mu a^\rho
-\bar D^\rho a^\mu\right)^c  \right\}.
\label{NA-Jfluc1}
\end{eqnarray}
\end{mathletters}%
For the fluctuations we find
\begin{mathletters}\label{NA-log-2}
\begin{eqnarray}
&&\left[v^\mu \bar D_\mu \,  \delta   {\cal J}^\rho \right]_a \, = 
-m^2_D v^\rho v^\mu 
\left[\bar D_\mu a_0-\bar D_0 a_\mu\right]^a \nonumber\\
&&\qquad\qquad\qquad\quad\,  -g f_{abc}  v^\mu a_\mu ^b \bar {\cal J}^{c,\rho} 
\ ,\label{vD-dJa} \\
&&\left[v^\mu \bar D_\mu  \delta {\cal J}^{\rho}\right]_{ab} =  g v^\mu a_\mu 
^m \left(
  f_{mac}\,\delta_{bd}
+f_{mbd}\delta_{ac}\right)   \bar{\cal J}^{\rho}_{cd}, 
\label{vD-dJab} \\
&&\left[ \bar D^2 a^\mu-\bar D^\mu(\bar D a)\right]_a+2 g f_{abc} 
\bar F_b^{\mu\nu}a_{c,\nu}=\delta J_a^\mu  \ .\label{dJa}
\end{eqnarray}
\end{mathletters}%

We solve the equations for the fluctuations (\ref{NA-log-2})
with an initial boundary condition for
$\delta f$, and $a_\mu(t=0)=0$. 
Exact solutions to (\ref{vD-dJa}) and (\ref{vD-dJab}) can be obtained 
\cite{future}. The current fluctuation $\delta {\cal J}_a$ reads, for 
$x_0 \equiv t \geq 0$,  
\begin{eqnarray}
\delta {\cal J}^\rho _a (x,v) &=& \bar U_{ab} (x, x_t) 
\, \delta {\cal J}^\rho _b (x_t, v)  \nonumber  \\
&-&    \int_0 ^{\infty}\!\!\!\! d \tau \, \bar U_{ab} (x, x_\tau)
\Big\{  g  f_{bdc}  v^\mu a_\mu ^d (x_\tau)
\bar {\cal J}^\rho_c(x_\tau, v)  
\nonumber \\
&& \quad\ +\ m^2_D v^\rho v^\mu \left(\bar D_\mu a_0
-\bar D_0 a_\mu\right)^b (x_\tau) \Big\} \,.\label{dJ-expl} 
\end{eqnarray}
We introduced  
$x_\tau \equiv x - v \tau$ and the 
parallel transporter  $\bar U_{ab}$, 
obeying   $v^\mu\bar D_\mu^x \, \bar U_{ab}(x,y)|_{y=x_\tau}=0$.
In order to solve (\ref{dJa})  for $a_\mu$, we make  a double expansion 
in both $\bar A$ and $\bar {\cal J}$,
using $\bar U_{ab} = \delta_{ab} 
+ {\cal O}(g \bar A)$. We denote by $a^{(n)}$ the term containing a 
total of $n$ powers in the mean fields $\bar A$ and/or $\bar {\cal J}$.

For our purposes, it will be sufficient to consider the zeroth 
order term in $\bar A$, and the zeroth and first order terms in 
$\bar {\cal J}$. Using the one-sided Fourier transform \cite{K}, we find
\begin{mathletters}\label{a+}
\begin{eqnarray}
a_{i,a\,+}^{T(0)}(k)&=&\frac{1}{-k^2+\Pi_T}
\int\!\!\frac{d\Omega_{\bf v}}{4\pi}\frac{\delta
{\cal J}^{T}_{i,a}(t=0,{\bf k},v)}{-i\ k \cdot v}\ , \\
a_{i,a\,+}^{T(1)}(k)&=&
\frac{-g f_{abc}}{-k^2+\Pi_T}P_{ij}^T({\bf k})
\int\frac{d\Omega_{\bf v}}{4\pi}\frac{1}{-i\ k\cdot v}\times\nonumber\\
&&\int\frac{d^4q}{(2\pi)^4} v^\mu a_{\mu}^{b(0)}(q) \ \bar{\cal J}_j^c(k-q,v)
\end{eqnarray}
\end{mathletters}%
in the gauge ${\bf k} \cdot {\bf a} =0$.
The function $\Pi_{T}(k)$ is the transverse polarization tensor of the 
plasma, $P_{ij}^T({\bf k})=\delta_{ij}-{k_ik_j}/{{\bf k}^2}$ the 
transverse projector, 
and $a_i^T\equiv P^T_{ij}a_j$.
Retarded boundary conditions are assumed above, with the prescription $k_0 
\rightarrow k_0 + i 0^+$.

With the above, we can express all fluctuations in terms of initial 
conditions and the mean fields. Following 
\cite{K} the statistical average over initial conditions can be deduced \cite{future}
and finally expressed (for each species or helicity index) as
\begin{eqnarray}
&& \langle\delta f_{{\bf k} pQ}\ \delta f_{{\bf k}' p' Q'}\rangle 
=(2\pi)^6\, \delta^{(3)}({\bf k}+{\bf k}')\times \nonumber\\
&&\qquad\left[\delta^{(3)}({\bf p}-{\bf p}') \delta (Q- Q')\, 
\bar f(p)
+\mu_{{\bf k} pp'QQ'}\right]\ .  \label{average}
\end{eqnarray}
The function $\delta (Q -Q')$  is, apart from a (representation dependent) 
normalization constant, a $N^2-N$ dimensional $\delta$-function 
over the proper set of Darboux variables related to the color charges 
\cite{KLLM}. The second term in (\ref{average}) is the Fourier 
transform of a smooth function
that vanishes at large distances. The above statistical average is all  
we need to evaluate the collision integrals.

For the remaining part we will concentrate on the dynamics of mean fields 
with typical momenta around $g m_D$. When computing the related 
collision integrals, we will find logarithmic divergences, cut-off in 
the infrared by the inverse collision time. We employ the leading logarithmic 
approximation, assuming $\ln(1/g)\gg 1$ while neglecting all sub-leading 
(though finite) terms.

We  find that the induced current $\langle J_{\mbox{\tiny fluc}}^{(0)}\rangle$
vanishes, as do the fluctuation integrals 
$\langle \eta^{(0)}\rangle$ 
and $\langle \xi^{(0)}\rangle$. The vanishing of $\langle \eta^{(0)}\rangle$
is consistent with the fact 
that in the Abelian limit the counterpart of  $\langle \eta\rangle$ vanishes 
at equilibrium \cite{K}. 
In the same spirit we evaluate the collision integrals containing one 
$\bar{\cal J}$ field. Consider
\begin{eqnarray}
&&\left\langle\xi^{(1)}_{\rho,a}\right\rangle 
= g f_{abc} v^\mu \left\{ -\left\langle   
a^{(1)}_{\mu,b} (x) \, \delta {\cal J}_{\rho,c}^{(0)}(x,v) 
\right\rangle \right. 
\nonumber\\
&&\quad +g f_{cde} v^\nu  \int^{\infty} _{0} \!\!\! d \tau 
 \bar{\cal J}_{\rho,e}( x_\tau, v) \left. \left\langle   a^{(0)}_{\mu,b} (x)\ 
a^{(0)}_{\nu,d} (x_\tau) \right\rangle \right\},
\end{eqnarray}
which simplifies, at logarithmic accuracy, to
\begin{eqnarray}
\label{leading-log}
\left\langle \xi^{(1)}_{\rho,a}(x,v)\right\rangle &=&
 -  \frac{g^2}{4\pi} N T \ln\left(1/g \right) 
\times \nonumber \\ && 
\qquad v_\rho\!\int\!
\frac{d\Omega_{{\bf v}'}}{4\pi}\, {\cal I}(v,v')\bar{\cal J}^0_a(x,v'), \\
{\cal I}(v,v') &\equiv&  \delta^{(2)}({\bf v}-{\bf v}') -\frac{4}{\pi}
\frac{({\bf v}\cdot{\bf v}')^2}{\sqrt{1-({\bf v}\cdot{\bf v}')^2}} \ .
\end{eqnarray}
The above expression has been 
obtained first in \cite{Bodeker}, and reproduces the collision integral 
considered in the Boltzmann equation of \cite{Arnold}.

We verified that the leading logarithmic solution is consistent with gauge 
invariance. Evaluating the correlator in (\ref{currentconservation}) yields 
$\bar D_\mu \bar J^\mu =0$, in accordance with (\ref{soft-YM}) in the present 
approximation.

Following B\"odeker, one can now estimate 
${\bar J}^i$ from (\ref{soft-mean}) to obtain for (\ref{soft-YM})
\begin{equation}
(\bar D_\mu {\bar F}^{\mu i})_a = \sigma \bar E^i _a + \nu^i_a  \ ,
\quad \sigma =\frac{4 \pi m_D^2}{3Ng^2T \ln \left(1/g\right)}\ .
\label{BEQ}
\end{equation}
This is the result of \cite{Bodeker}. The coefficient 
$\sigma$ represents the color conductivity and has been discussed 
in \cite{Arnold,Selikhov2}. The white noise $\nu$ 
has its origin in the fluctuations of the transverse part of $\xi^{(0)}$ 
\cite{Bodeker,future}. We obtain to leading order
\begin{equation}\label{noise}
\left\langle \nu_{a}^{i}(x)\ \nu_{b}^{j }(y)\right\rangle =  
2\, T \,\sigma\, \delta^{ij}\delta_{ab}\, \delta^{(4)}(x-y)\ ,
\end{equation}
in accordance with the fluctuation-dissipation theorem (FDT).
Note also that the classical Debye mass differs from the quantum one.

In order to go beyond classical transport theory we expand about
the bosonic (fermionic) quantum-statistical equilibrium distribution function
$\bar f_+$ ($\bar f_-$). For gluons in the adjoint, the Debye mass obtains as 
$m_D^2 = g^2 NT^2/3$. The FDT is obeyed as well, if ${\bar f}$ in 
(\ref{average}) is replaced by ${\bar f_\pm} (1 \pm {\bar f_\pm})$. 
(This should however be derived in a similar way as (\ref{average}) 
from the microscopic theory \cite{future}.)
Also, the quantum collision integrals are obtained with the correct 
statistical factors \cite{future}. 
It is interesting to note that all quantum 
modifications are contained in the implicit change of $m_D$. 

This terminates the explicit derivation, 
in the leading logarithmic approximation, of the collision integral 
and the dynamical equations for the soft fields 
from classical transport theory.

Summarizing, we have given a prescription to derive mean gauge
field equations from classical transport theory. This includes a recipe
to obtain effective (classical or quantum)
collision integrals from the microscopic theory.
The approach is in accordance with gauge invariance. 
In a close-to-equilibrium plasma and for small gauge coupling, 
we reproduce B\"odeker's effective theory.

The last part of our analysis can straightforwardly be generalized in order
to obtain explicit expressions for the collision integrals not only for
the soft momentum region.
Another interesting open problem is using the same methods 
for out-of-equilibrium 
situations. Based on the evaluation of collision
integrals for Abelian plasmas  out of equilibrium \cite{K},
we should find the Coulomb logarithm changing drastically 
the mean non-Abelian gauge field equations. 

It remains remarkable that classical transport theory is efficient 
enough as to describe not only the non-Abelian dynamics of semi-hard modes
with momenta around $m_D$, but as well the non-perturbative dynamics 
of soft gluons at leading logarithmic order.
This establishes a link even beyond the one-loop level between our 
approach and a complete quantum field theoretical treatment, 
whose deeper structure is waiting for being uncovered \cite{Pisarski}.


\begin{thebibliography}{99}
\bibitem{ElzeHeinz}H.-Th.~Elze and U.~Heinz,  Phys.~Rep.~{\bf 189}, 81 (1989).
\bibitem{Gross}
D.~J.~Gross, R.~D.~Pisarski, and L.~G.~Yaffe, Rev. Mod. Phys. 
{\bf 53}, 43 (1981).
\bibitem{Bodeker}
D.~B\"odeker, Phys. Lett. {\bf B426}, 351 (1998).
\bibitem{Heinz}
 U.~Heinz, Phys. Rev. Lett. {\bf 51}, 351 (1983); Ann. Phys. 
{\bf 161}, 48 (1985); ibid {\bf 168}, 148 (1986).
 \bibitem{KLLM}
P.R. Kelly, Q. Liu, C. Lucchesi and C. Manuel, Phys. Rev. Lett {\bf 72}, 
3461 (1994); Phys. Rev.
{\bf D 50}, 4209 (1994).
\bibitem{HTL} 
R.~D.~Pisarski, Phys. Rev. Lett {\bf 63}, 1129 (1989); E.~Braaten and  
R.~D.~Pisarski,
Nucl. Phys. {\bf B337}, 569 (1990); J.~P.~Blaizot and E.~Iancu,   
Phys. Rev. Lett {\bf 70}, 3376
(1993); Nucl. Phys. {\bf B417}, 608 (1994).
\bibitem{future}D.F. Litim and C. Manuel, {\tt hep-ph/9906210}.
\bibitem{K}
Yu. L. Klimontovich, ``{\it Statistical Physics}", Harwood Academic 
Publishers, (1986); E.~Lifshitz and 
L.~Pitaevskii, ``{\it Physical Kinetics}", Pergamon Press, Oxford (1981).
\bibitem{Wong}
S. Wong, Nuovo Cim. {\bf 65A}, 689 (1970).
\bibitem{BGS}L.F. Abbott, Nucl.~Phys.~{\bf B185}, 189 (1981); 
H.-Th.~Elze, Z.~Phys.~{\bf C47}, 647 (1990).
\bibitem{Selikhov}A.V. Selikhov, Phys.~Lett.~{\bf B268}, 263 (1991), 
Erratum  Phys.~Lett.~{\bf B285}, 398 (1992).
\bibitem{SG}
A.V. Selikhov and M. Gyulassy, Phys. Rev. {\bf C 49}, 1726 (1994).
\bibitem{Arnold}
P.~Arnold, D.~T.~Son, and  L.~G.~Yaffe, hep-ph/9810216.
\bibitem{Selikhov2}A.V. Selikhov and M. Gyulassy, Phys.~Lett.~{\bf B316}, 
373  (1993); H.~Heiselberg, Phys. Rev. Lett. {\bf 72}, 3013 (1994).
\bibitem{Pisarski}R.~D. Pisarski, hep-ph/9710370. 

\end{thebibliography}
\end{document}